%% file: main.tex
\documentclass[sigconf, nonacm]{acmart}

\usepackage{listings}
\usepackage{xcolor}
\usepackage{xspace}
\usepackage{soul}
\usepackage{colortbl}
\usepackage{enumitem} 
\usepackage{csquotes}
\usepackage{marginnote}
\usepackage{multirow}
\usepackage{graphicx}
\usepackage{booktabs}
\usepackage[section]{placeins}
\usepackage{xurl}
\usepackage{subfigure}
\usepackage{seqsplit}
\usepackage{comment}
\usepackage{algorithm}
\usepackage{algpseudocode}
\usepackage{sidecap}
\usepackage{balance}

\sidecaptionvpos{figure}{c}

\usepackage{dialogue}

\newtheorem{definition}{Definition}

\usepackage{mdframed}
\newmdenv[backgroundcolor=blue!10, hidealllines=true, linewidth=\linewidth]{mychatusercolor}
\newmdenv[backgroundcolor=green!10, hidealllines=true, linewidth=\linewidth]{mychatmodelcolor}

\usepackage{spverbatim}
\newcommand{\tinyskip}{\vspace{3pt}}
\newcommand{\mypar}[1]{\tinyskip\noindent\textbf{#1.}\xspace}

\newenvironment{myitemize}{%
\begin{itemize}[leftmargin=1em, itemsep=.1em, parsep=.1em, topsep=.1em,
    partopsep=.1em]}
{\end{itemize}}

\newenvironment{myenumerate}{%
\begin{enumerate}[leftmargin=1em, itemsep=.1em, parsep=.1em, topsep=.1em,
    partopsep=.1em]}
{\end{enumerate}}

\newenvironment{structure*}{\color{blue}\begin{myenumerate}}{\end{myenumerate}}

\sethlcolor{yellow}

\newcommand{\nameofsys}{DePLOI}

\begin{document}
\title{\nameofsys: Applying NL2SQL to Synthesize and Audit Database Access Control}

\author{Pranav Subramaniam$^1$, Sanjay Krishnan$^1$}
\affiliation{
\institution{$^1$ University of Chicago}
}
\email{{psubramaniam, skr}@uchicago.edu}

\begin{abstract}
In every enterprise database, administrators must define an access control policy that specifies which users have access to which tables. Access control straddles two worlds: policy (organization-level principles that define who ``should'' have access) and process (database-level primitives that actually implement the policy). Assessing and enforcing process compliance with a policy is a manual and ad-hoc task. This paper introduces a new access control model called \emph{Intent-Based Access Control for Databases} (IBAC-DB). In IBAC-DB, access control policies are expressed using abstractions that scale to high numbers of database objects, and are traceable with respect to implementations. This paper proposes \nameofsys\ (\underline{De}ployment \underline{P}olicy \underline{L}inter for \underline{O}rganization \underline{I}ntents), a LLM-backed system leveraging access control-specific task decompositions to accurately synthesize and audit access control implementation from IBAC-DB abstractions.
As \nameofsys\ is the first system of its kind to our knowledge, this paper further proposes IBACBench, the first benchmark for evaluating the synthesis and auditing capabilities of \nameofsys. IBACBench leverages a combination of current NL2SQL benchmarks, real-world role hierarchies and access control policies, and LLM-generated data. We find that \nameofsys\ achieves high synthesis accuracies and auditing F1 scores overall, and greatly outperforms other LLM prompting strategies (e.g., by 10 F1 points).
\end{abstract}

\maketitle

\input{sections/intro}
\input{sections/bg}
\input{sections/method}
\input{sections/benchmark}
\input{sections/eval}
\input{sections/conclusion}
\newpage
\balance


\bibliographystyle{ACM-Reference-Format}
\bibliography{sample}

\end{document}

%% file: sections/intro.tex
\section{Introduction}
\begin{figure}
  \centering
  \includegraphics[width=0.9\columnwidth]{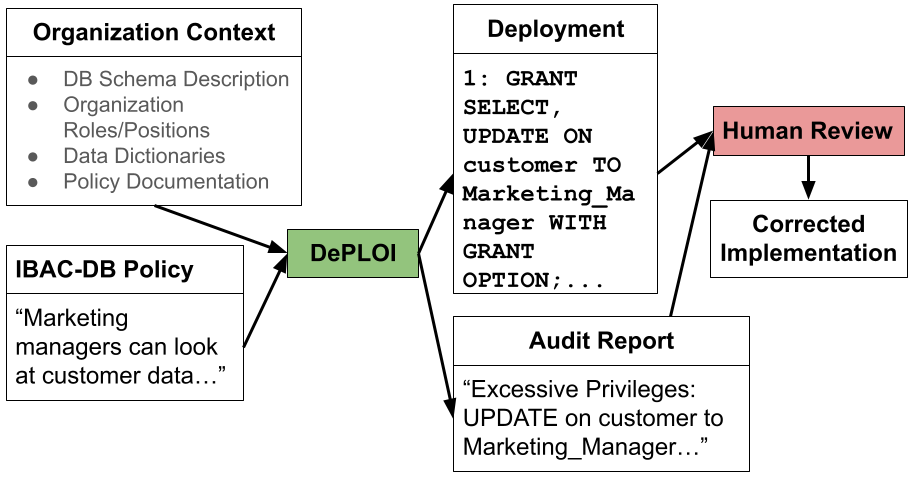}
  \caption{A Workflow for Using LLM-based Systems to Synthesize and Audit Database Deployments} 
\label{fig:workflow}
\end{figure}

Database deployment is notoriously complex, and the DB community has a long history of automating various aspects of deployment, ranging from learned indexes to self-driving databases to NL interfaces for querying.
However, deployments increasingly have to comply with internal and external policies around security, privacy, and data quality (e.g., requirements such as "only senior engineers can have write access to production databases").

A key aspect of these policies is \emph{database access control}, which specifies, among other properties, \emph{access control permissions}, the permitted SQL operators for roles/users on database tables/views, \emph{temporal constraints}, the times during which a role/user is allowed to access a database view, or the \emph{role hierarchy}, which indicates which roles inherit permitted SQL operators from other roles.

Configuring the various aspects of database access control straddles two different worlds: policy and process, which requires matching text to database components. Currently, performing this matching requires tedious manual communication between legal experts who write policy and database experts who write process.

With the rise of large language models (LLMs) capable of sophisticated language understanding and code synthesis, it may be possible to automatically derive fragments of process implementations directly from policy documents, reducing this effort.
However, in the case of access control, there are several challenges: (i) policy documents specify multiple types of complex interdependent access control requirements. This requires synthesizing SQL programs consisting of \emph{multiple interdependent queries} from a \emph{large policy document}, contrary to the existing NL2SQL setting, which only considers the translation of a single question to a single query.
(ii) Based on state-of-the-art NL2SQL solutions, it is unrealistic to expect that synthesizing such SQL programs will be accurate. The scale of input NL and output SQL we must generate is much larger than in the typical NL2SQL setting. Further, certain access control requirements may not be enforced at all times, potentially leading to security risks (e.g., a data analyst may temporarily require access to a view they normally cannot access, and a DB admin may simply forget to revoke access privileges once the analyst has finished their task). 
Therefore, it is important for downstream error-correcting workflows (e.g., debugging, etc.) to be able to \emph{trace} synthesized implementation to specific parts of a policy, and \emph{audit} implementations to help determine which parts of implementation may be incorrect.
(iii) Current NL2SQL benchmarks do not consider the task of auditing, and do not consider the domain of access control. Therefore, we require a novel benchmark design for synthesis and auditing access control policy implementations.

\begin{figure}
  \centering
  \includegraphics[width=0.9\columnwidth]{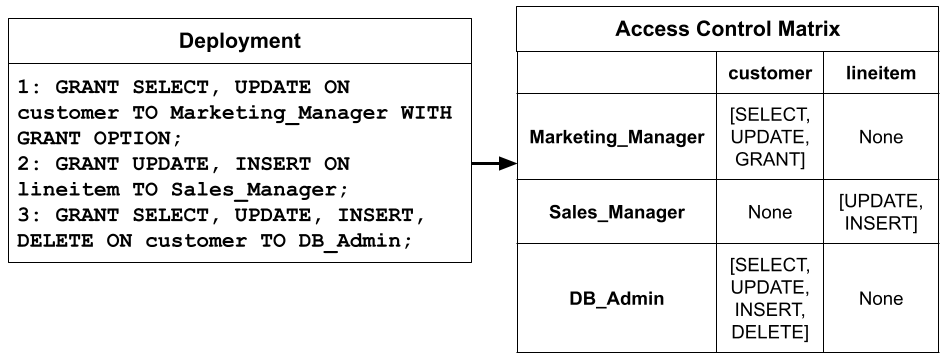}
  \caption{From SQL GRANT statements to Access Control Matrix. Roles are row indexes, tables/views are column headers, and permitted operations are matrix cells.} 
\label{fig:sql2acm}
\end{figure}

To address the scale and complexity of NL and SQL (challenge (i)) and enable the implementation to be traceable to the policy (challenge (ii)), we observe the following: access control matrices (ACMs), current abstractions leveraged by access control models, can easily represent any number of roles and tables/views. And they are easily traceable from an implementation: given a set of SQL GRANT statements, one can simply rearrange the roles, views, and permissions into rows, column headers, and cells of a matrix, and this forms the Access Control Matrix describing a database (see Figure~\ref{fig:sql2acm}).
Based on this, instead of expressing access control policy requirements using documents, we propose a novel access control model that \emph{permits expression of access control using natural language} unlike existing access control models. We call this \emph{Intent-based Access Control for Databases}, or IBAC-DB. 
The IBAC-DB model represents various access control policy requirements and their implementations using a novel set of ACM abstractions that allow natural language as matrix elements.

We propose \nameofsys\ (\underline{De}ployment \underline{P}olicy \underline{L}inter for \underline{O}rganization \underline{I}ntents), a LLM-backed system \emph{which compares a database deployment to an intended configuration expressed using IBAC-DB}. Much in the same way that software engineering increasingly uses sophisticated linting checks to catch potential bugs, we envision a framework that can audit database deployments for potential misconfigurations.
\nameofsys\ could be used in a workflow such as the one shown in Figure~\ref{fig:workflow} in which humans create IBAC-DB policies and organizational context (such as metadata) to be given as input to an LLM-backed method, and humans review and correct the output deployment. A LLM-backed method in such a workflow can be used to speed up implementation synthesis and audit implementations for potential misconfigurations with respect to policy.

\nameofsys\ leverages the key insight that the learning task of policy synthesis and auditing can be decomposed into the following subtasks that can be accurately performed by LLMs: \emph{role-view mapping}, \emph{SQL operator comparison}, and \emph{Temporal constraint comparison}, and each of these subtasks can be further decomposed into combinations of program comprehension and condition comparison subtasks to be used in a least-to-most prompting strategy that we find is highly accurate.
The interdependencies among these subtasks can then be accounted for in synthesizing and auditing implementation via a simple procedure.

To evaluate \nameofsys, we construct \textbf{IBACBench}, the first benchmark of its kind, for testing synthesis and auditing systems for implementing access control policies. IBACBench consists of access control policies, both synthetically constructed using LLMs, and real-world policies over the Sociome database and role hierarchies from the Amazon Access dataset. We also leverage existing NL2SQL benchmarks for databases and views over which to enforce access control.
We further observe that the evaluation metric used by current NL2SQL metrics, Test Suite Accuracy~\cite{zhong-etal-2020-semantic}, ignores semantically equivalent but tuple-inequivalent answers, where extra attributes are returned with the correct tuples. We propose a novel evaluation metric, \emph{Attribute Subset Execution Accuracy}, which counts such cases as correct as well. Under this metric for synthesis, and F1 scores for auditing, we find that \nameofsys\ is highly accurate on IBACBench compared to alternative prompting strategies.
To our knowledge, \nameofsys\ is the first system for synthesizing and auditing access control. However, using the performance of state-of-the-art NL2SQL systems on current NL2SQL benchmarks as reference, we further observe that \nameofsys's accuracy is similar or better.

\subsection{Policy Linting Concerns}
\noindent \textbf{Why LLMs?} There is a lot of context that must be considered when implementing database deployments, and this context can greatly vary in format and content. LLMs are polymorphic: they can take as input any context as long as context can be turned into a string.

\noindent \textbf{But LLMs Hallucinate.} Yes, using LLM outputs simply as-is would not be acceptable. However, LLMs can help reduce the effort needed and risk of error in implementing database deployments from policy by speeding up deployment code synthesis and auditing the results for errors, meaning end users have less work--no need for writing code from scratch, or examining the whole codebase for potential errors. This is why we propose using a LLM-backed method such as \nameofsys\ in a workflow such as Figure~\ref{fig:workflow}.

\noindent \textbf{Why is this relevant to the DB community?} Many aspects of DBMS configuration have been improved through AI, including performance tuning (e.g., AI algorithms for query optimization~\cite{marcus2019neo}), physical design (e.g., learned indexes~\cite{kraska2019sage}), and improved database interfaces using NL2SQL~\cite{floratou2024nl2sql, catsql2023}. However, automating database deployment has received little attention. We believe there is a rich set of research challenges required to achieve this that are relevant to the DB community.

\noindent \textbf{How do I know that \nameofsys\ is accurate?} \nameofsys\ is the first system that synthesizes and audits implementation for interdependent access control requirements.
As such, there is not yet another state-of-the-art system to compare against. That said, we evaluate \nameofsys\ with respect to alternative prompting strategies that use various other forms of reasoning and task decomposition, and we find \nameofsys\ is far more accurate in comparison.
Further, we note that performing traditional NL2SQL is required to determine the data over which to define access control rules.
Therefore, we use accuracies of state-of-the-art systems for reference-namely, as an upper bound on performance.
Notably, the performance of the state-of-the-art systems on many NL2SQL benchmarks is far from perfect--for example, on the BIRD benchmark, the highest performance observed is 77.3\%~\cite{birdleaderboard}. On the Spider benchmark, the highest performance is 94\%~\cite{yu-etal-2018-spider}.
We find that with respect to these numbers, \nameofsys's synthesis performance is on par with state-of-the-art NL2SQL systems.

%% file: sections/bg.tex
\section{Background}
\subsection{Existing NL2SQL Capabilities}
\label{subsec:existing}
NL2SQL systems seem helpful in implementing NL access control policies as SQL. 
Current NL2SQL capabilities focus on converting natural language questions about information in a database to queries (typically SELECT queries)~\cite{dong2023c3, yu-etal-2018-spider, ni2023lever, wang-etal-2020-rat}. The state-of-the-art NL2SQL methods develop procedures for prompting LLMs to generate accurate SQL, such as ChatGPT~\cite{dong2023c3} or GPT-4~\cite{gao2023texttosql}.

However, to our knowledge, current NL2SQL benchmarks do not contain examples that generate access control queries~\cite{yu-etal-2018-spider, zhongSeq2SQL2017, fiben2020}, and the database system used for evaluation in most cases is SQLite, which does not support access control.

Further, it is not obvious how to bridge this gap between state-of-the-art NL2SQL methods and the access control use case. For example, enhancing LLM-backed NL2SQL systems by adding naive prompting of ChatGPT can produce an incorrect answer. When given the prompt, \emph{Write the postgresql commands to implement such access control, given the database schema}, two of the queries generated by ChatGPT are "GRANT CREATE ON ALL TABLES IN SCHEMA public TO data\_architect;" and
"GRANT CREATE ON ALL VIEWS IN SCHEMA public TO data\_architect;".
The second query is unnecessary, and uses the word "VIEWS" in the grant statement which does not exist.

Simply using NL2SQL methods as-is by inputting access control policy documents can also be inaccurate. Consider a straightforward access control rule: "Grant the user John select access on the customer table with the option of passing down this privilege." C3~\cite{dong2023c3} translates this NL to the SQL "GRANT SELECT ON customer TO John;". This is incorrect because it ignores the option of passing down the privilege--the query must be suffixed with a "WITH GRANT OPTION".

\begin{figure}
  \centering
  \includegraphics[width=\columnwidth]{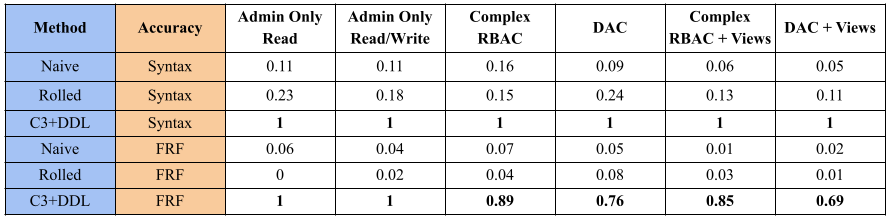}
  \caption{Comparison of NL2SQL Translation by LLM method, with respect to syntax and Forward-Reverse-Forward (FRF) Accuracy. Columns are access control policies.}
\label{fig:llmnl2sql}
\end{figure}

On the other hand, we observe that, when given a policy as a set of natural language sentences, each of which clearly spells out the role, table/view, and privilege, C3 can be altered slightly to very accurately generate GRANT statements, particularly compared to naive prompting, or prompts engineered using typical prompt engineering techniques. 
We observe this via the following initial experiment, whose results are shown in Figure~\ref{fig:llmnl2sql}: given a NL sentence that indicates the role, table/view and privilege to be granted, we run three different methods to generate the correct GRANT statement: (i) \emph{C3+DDL}: our new variant of C3 where we alter C3 by adding access control-specific rules and slightly changing its completion-style prompts (instead of ending in "SELECT" to generate a SELECT query, they end in GRANT, or CREATE VIEW, to define new views and grant privileges), (ii) \emph{Naive prompting}: we simply give ChatGPT the input sentence with the following instruction: "Write the postgresql commands to implement such access control, given the database schema." (iii) \emph{Rolled prompting}: we use chain-of-thought prompting, where the reasoning steps are: (1) first identify the parts of the sentences that may denote the roles, tables/views, and privileges (2) Write the commands to grant the privileges indicated by the sentence.
In Figure~\ref{fig:llmnl2sql}, we compare the Naive, Rolled, and C3+DDL methods with respect to accuracy of the generated syntax, (labeled "Syntax" in the figure) and forward-reverse-forward (labeled "FRF" in the figure) accuracy.
We systematically perform this comparison on access control policies containing role hierarchies of various complexities, view creation, and discretionary access control. We synthetically generate 100 policy documents of each type, and compute the accuracy based on the syntax/FRF accuracy of the synthesized SQL script. We see that C3+DDL's Syntax and FRF accuracy far exceeds prompting solutions.

Based on these examples and initial experiment, we hypothesize the following: if access control policies can be formatted to clearly reflect the roles, privileges, and views to implement on the database, this enables task decomposition: database roles, database views, and database privileges can be generated separately. For each of these subtasks, the prompting from NL2SQL methods can be adapted to accurately implement it, allowing for accurate access control implementation.
This method would also have much lower cost than gathering large amounts of training data and attempting to train an access control-specific NL2SQL system.

\subsection{A Case for Auditing}
Currently, NL2SQL techniques focus only on converting a NL question to a SQL query. Although NL2SQL techniques are far from accurate~\cite{birdleaderboard}, most NL2SQL systems do not include methods to audit output queries for potential inaccuracies. Instead, this is treated as a separate problem handled by ad-hoc methods such as simply debugging or coding assistants, for example~\cite{lakehouseiq, runai}.

In the case of access control, policies can contain larger amounts of text than just a single question, and the output of NL2SQL for access control is a program of multiple interdependent SQL statements, making such ad-hoc auditing tedious and error-prone.
An alternative in the access control literature is \emph{combinatorial testing}: (i) manually devising the only ground truth implementation, and then
(ii) combinatorially generating tests to compare whether allowed
operations under the ground truth are allowed under the system-generated
implementation, and vice versa~\cite{bertolino2018oracles, li2015evaluating, zhang2023trust}. These steps are also tedious.
To address this issue, we propose auditing implementation by using LLMs to \emph{directly compare} the NL policy to its SQL implementation.

In this paper, we propose a system that uses LLMs to both \emph{synthesize} access control SQL and \emph{audit} SQL with respect to policy. This will make the implementation of access control less tedious and error prone when used in a workflow like the following (shown in Figure~\ref{fig:workflow}): (1) Users will write access control policies (2) Access control policies will be inputted into the system (3) The system will synthesize SQL and audit the synthesized implementation with respect to the policy (4) The end user will receive the implementation and audit report, which they will use to correct implementation errors. 


\subsection{Access Control Related Work}
In this section, we cover key access control definitions and abstractions. We observe that current access control models do not permit NL, making it difficult to trace access control implementation to text in a policy.
\subsubsection{Access Control Definitions}
Let us consider access control to a single database. Over this database, there are a set of possible users $U$. These users can run a set of possible SQL statements $S$ over this database.
An access control \emph{scheme} is a set of rules that specifies who should be given access under what conditions. Formally, an access control scheme for a database is a program, that determines whether a particular user is allowed to run the desired SQL statement over the database:
\[
\mathcal{P}(user, stmt): U \times S \mapsto \{allowed, denied\}
\]

$\mathcal{P}$ describes a literal implementation of access control for a particular database.
For example, in an SQL database, $\mathcal{P}$ is defined as a sequence of GRANT statements. However, such programs are not created in a vacuum. An access control \emph{policy} is a set of rules that specifies who should be given access to what resources. For example, an organization's regulatory posture may dictate who has access to data (e.g., only relevant doctors may view electronic health records).
Or, an organization's security team may implement a particular minimal privilege strategy (e.g., only senior management can modify customer records). Whatever the reason, these decisions are described at a higher level of abstraction than $\mathcal{P}$.

There are many formats in which access control policies may appear. SOC-2 compliance requires that access control policies be specified in an access control matrix~\cite{aicpa}. 
An access control matrix defines rows that are users (or groups) and columns that are data assets. The cells of this matrix specify the allowed operations of that user on the asset.
As another example, research institutions that work with identifiable information need to submit data management plans written in document form to institutional review boards and funding agencies~\cite{loyolaacp, nwpacp}. 

In general, access control policies are written in natural language or quasi-natural language (e.g., a document containing loosely written access control matrices) by a database non-expert. To correctly implement such policies, one must determine which parts of the policy represent access control rules for the database, and to which components of the database those parts refer.
This can require multiple rounds of communication between a policy expert and a database expert (e.g., a database administrator)~\cite{oreillych3, nistacguide, satoriacsteps, actemplate}. 

\subsubsection{Auditable Access Control}
Access control implementations must be audited to determine whether they comply with a given policy. This involves comparing two access control schemes $\mathcal{P}$, which is a scheme specified on the database in SQL (e.g., a list of GRANT statements), and $\mathcal{P}^*$, which is a scheme specified in natural language (e.g., an organization's access control policy document). Hereafter, we will refer to $\mathcal{P}$ as an access control ``implementation'' and $\mathcal{P}^*$ as an access control ``policy''. We fully understand that there is inherent imprecision in natural language and there might not exist an unambiguous $\mathcal{P}^*$; however, we scope our initial exploration of this problem in such a way to avoid such ambiguities. The main goal is to be able to check for compliance, or formally:

\begin{definition}[Compliance]
An access control implementation $\mathcal{P}$ is compliant against a policy $\mathcal{P}^*$ if and only if \[~~\forall (u,t) \in U \times T: ~~ \mathcal{P}(u,t) \implies \mathcal{P}^*(u,t)\] 
\end{definition}

This definition states that every allowed SQL statement in $\mathcal{P}$ is also allowed in  $\mathcal{P}^*$, or in other words, $\mathcal{P}$ is at least as restrictive $\mathcal{P}^*$.
Many solutions have been developed that enable automated implementation and verification of access control policies. However, we will see that these come at the cost of restricting access control rules that can be expressed.

One solution is to adhere to an access control paradigm for specifying access control policies, such as role-based access control (RBAC), discretionary access control (DAC), etc. Adhering to these paradigms alone can greatly simplify defining and implementing access control policies. For example, if one defines a role-based access control policy on a database, one need only implement it by concatenating the SQL describing the roles, tables/views, and privileges into a SQL GRANT statement. However, implementing such policies through a database system alone requires database expertise and explicit enumeration of the complete set of access control rules for all database users. 

To reduce the expertise and manual effort needed to implement an access control policy, prior work has focused on allowing access control to be defined via programming languages for access control~\cite{nistac2017, shilldb2019, xacml2003}, which can then be translated into database privileges, often with guarantees on the correctness of translation, or automated verification via model-checking. 

These programming languages allow automatic and provably correct implementation of access control policy. However, they restrict the access control rules that can be expressed, compared to natural language.
For example, ShillDB, a recent contract language for database access control, cannot express access control rules on nested queries~\cite{shilldb2019}, but natural language could easily express access control rules on such a query (e.g., Q4, the order priority checking query, from TPC-H).

Based on existing access control solutions, we find that on the one hand, one can use NL to represent policies, but then implementation and auditing for correctness are manual. On the other hand, you can use access control paradigms or languages which automatically and correctly implement access control policies, but then the expression of access control rules is limited.

\subsection{Problem Statement}
In this paper, we solve the following problem: let $R$ be the set of roles in an organization, let $P$ be a policy document describing access controls over a database with schema $\mathcal{S}$ (we assume $\mathcal{S}$ is a set that contains both database tables and user-defined views). Then, $\forall r \in R$, policy $P$ describes three types of access control requirements: (i) \emph{access control permissions}: database operators permitted on tables or views of the database. Formally, let $\mathcal{O}$ be the set of permitted database operators (e.g., SELECT, UPDATE, etc.). Then, the set of access control permissions for role $r$ is: $P'_r = \{ (O, v) | O \subset \mathcal{O}, v \in \mathcal{S} \}$. 
(ii) \emph{role inheritances} $\mathcal{R}'_r \subset \mathcal{R}$, where $r' \in \mathcal{R}'_r$ means that $r$ will have all access control permissions that $r'$ has.
(iii) \emph{temporal constraints}, which represent the times during which role $r$ can perform certain database operations on certain tables/views. Formally, let $\mathcal{T_P}$ be the set of all \emph{periodic expressions} (informally, a set of time intervals-formal definition in ~\cite{ferrari2000trbac}). Then, the set of temporal constraints for $r$ is $\mathcal{C'_r} = \{ (T, O, v) | T \in \mathcal{T_P}, O \subset \mathcal{O}, v \in \mathcal{S} \}$, where each triple $(T, O, v)$ represents $T$, a periodic expression, $O$, the set of database operators permitted during $T$, and $v$, the database table/view over which the permissions and periodic expression apply.

Then, we solve the following problem:

\begin{myenumerate}
    \item \textbf{Synthesis}: Given NL policy document $P$, \emph{synthesize} a sequence of queries $Q$ that implements these access control permissions $P'_r$ and role inheritances $\mathcal{R}'_r$ using the correct GRANT statements, and temporal constraints $\mathcal{C'_r}$ using the correct row-level policy statements (CREATE POLICY statements).
    \item \textbf{Auditing}: Given NL policy document $P$ and a sequence of queries, $Q$, \emph{audit} $Q$ to check if it complies with $P$.
\end{myenumerate}

Solving the Synthesis problem requires addressing several challenges: (i) \emph{NL2SQL}: all challenges associated with the typical NL2SQL setting must be considered here in order to correctly synthesize SQL queries representing database views. Namely, we must consider the \emph{schema linking} challenge, particularly in the case where schema elements are domain-specific, heavily abbreviated, and require external information specific to an organization to successfully link schema elements to text.
(ii) \emph{Permission Mapping}: we must map allowed actions denoted by text to SQL operators. (iii) \emph{Temporal Constraint Mapping}: we must map text related to time intervals to SQL conditions.
(iv) \emph{Interdependent Requirements}: synthesizing individual queries for permissions, temporal constraints, and inheritances is not enough--the queries must be in the \emph{correct order}. 
Formally, let $Q$ be a sequence of SQL statements that implement access control in policy $P$. 
Then, for every pair of queries in $Q$, $q_i$, $q_j$, $i < j$, such that both $q_i$ and $q_j$ specify access control over the same database role and view, we say $q_i$ and $q_j$ are in the correct order if $q_j$ does not overwrite the conditions on permissions assigned by $q_i$. 

\noindent\textbf{Interdependence Example.} Suppose $q_i$ is an inheritance condition, e.g., "GRANT software\_engineer TO vp\_of\_engineering", and that the role "software\_engineer" can SELECT and INSERT on the customer table.
Suppose $q_j$ is a grant statement, "GRANT SELECT, INSERT, UPDATE ON customer TO vp\_of\_engineering".
Then, $q_j$ could overwrite $q_i$'s condition if it grants permissions that were supposed to be inherited, resulting in incorrect permissions later on.
For example, if "SELECT" permissions are revoked from "software\_engineer", they will not be revoked from "vp\_of\_engineering" if $q_j$ appears after $q_i$, even though "vp\_of\_engineering" is supposed to inherit SELECT privileges from "software\_engineer".

As another example, suppose $q_i$ grants SELECT access only from 9am to 5pm on weekdays (Monday through Friday). Suppose $q_j$ grants SELECT access on the same role and same view (e.g., "GRANT SELECT ON customer TO software\_engineer").
Then, if $q_j$ appears after $q_i$, then $q_j$ will overwrite $q_i$ because it grants SELECT without any temporal condition.

Solving the Auditing problem requires addressing several challenges: (i) \emph{traceability}: a downstream workflow for correcting errors, such as Figure~\ref{fig:workflow}, should be able to trace synthesized implementation to specific text in a policy document. (ii) \emph{auditing procedure}: we require a novel procedure to audit implementation to determine if it complies with policy.

\noindent\textbf{Problem Scope.} We assume there is \emph{data scarcity}--that is, we do not have enough data to train ML pipelines, or fine-tune LLMs. Therefore, in this paper, the space of our solutions will consist of prompting workflows. We believe this assumption is justified because realistically, organizations possess a very small number of security policy documents in which access control is described, and the formats and training domains in which access control is described can vary greatly across organizations.

To address these challenges, we propose a novel access control model for expressing access control requirements in natural language, \emph{Intent-based Access Control for Databases} (IBAC-DB). We propose \nameofsys, a system that synthesizes and audits policies expressed using IBAC-DB. Concretely, we make the following contributions:
\begin{myenumerate}
    \item \textbf{Traceability Solution}: We propose a new paradigm for access control called \emph{Intent-based Access Control for Databases} (IBAC-DB), which uses the \emph{natural language access control matrix} (NLACM) as the input.
    \item \textbf{Synthesis Solution}: \nameofsys\ takes IBAC-DB abstractions as input and accurately synthesizes access control implementation. \nameofsys\ achieves this by leveraging these abstractions' structures to decompose the task of synthesis into the subtasks of synthesizing roles, views, and the different types of privileges separately.
    \item \textbf{Auditing Solution}: To allow policy auditing, we define \emph{differencing}, a procedure to compare an access control policy and its implementation. Differencing similarly leverages task decomposition to the subtasks: \emph{role-view mapping} to compare roles and views, and \emph{privilege comparison}, which is carried out using a least-to-most prompting strategy.
    Note that there is no way to formally verify the correctness of implementations, due to the imprecision of natural language. However, we find that our differencing procedure for auditing policies is mostly correct in practice.
\end{myenumerate}

%% file: sections/method.tex
\section{\nameofsys\ Overview}
We first describe the novel access control model we use to represent both policy documents and SQL implementations.
Then, we present \nameofsys, our LLM-backed system for synthesizing and auditing access control policy implementations.

\subsection{IBAC-DB}
Database access control rules are specified as GRANT statements, each of which consists of the allowed SQL operators for a role/user on a table/view.
We refer to NL or SQL that references key elements of access control rules (roles/users, tables/views, permitted SQL operations, etc.) as \emph{intents}. Then, intent-based access control (IBAC-DB) enables the specification of access control policies using intents. IBAC-DB achieves this using several novel abstractions: the \emph{natural language access control matrix}, or NLACM, the \emph{temporal access control matrix} or TACM, and the \emph{role hierarchy list}, or RHL.

A NLACM specifies one database access control permission per matrix cell: a row represents privileges for a database role/user, a column represents privileges for a table/view, and a cell represents the allowed SQL operators.
A TACM specifies one temporal access constraint per matrix cell: a row represents privileges for a database role/user, a column represents privileges for a table/view, and a cell represents a temporal interval during which access is permitted for the role/user on the table/view.
A RHL specifies one role inheritance relationship per list element: a row represents a parent and child role, where the child inherits all access control permissions granted to the parent. Formally,

\mypar{Definition of NLACM}: Let $D$ be a database schema, which consists of: (i) table schema definitions, (ii) view definitions, (iii) role and user definitions. Then, a NLACM is a $m \times n$ matrix where each row represents the database privileges of a role/user in the database, and each column represents a table/view. The $(i, j)$th cell represents the permitted SQL operators of role $i$ on table/view $j$.

\mypar{Definition of TACM}: Let $D$ be a database schema, which consists of: (i) table schema definitions, (ii) view definitions, (iii) role and user definitions. Then, a TACM is a $m \times n$ matrix where each row represents the database privileges of a role/user in the database, and each column represents a table/view. The $(i, j)$th cell represents a periodic expression denoting the times during which role $i$ can access table/view $j$.

\mypar{Definition of RHL}: Let $D$ be a database schema, which consists of: (i) table schema definitions, (ii) view definitions, (iii) role and user definitions. Then, a RHL is a list of pairs of size $m$ where each pair represents a parent role/user and a child role/user in the database.

These IBAC-DB abstractions specify access control rules not already implemented in the database. Toward this, they have the following constraints:
\begin{myitemize}
    \item Each role/user appears in only one row/element.
    \item Each view appears in only one column of a NLACM or TACM.
    \item (Principle of Failsafe Defaults): There can be roles, views in the DB that do not appear in these abstractions, and we assume they do NOT have any privileges defined.
    \item Empty cells are permitted, indicating no privilege is assigned.
    \item Non-empty cells indicate which SQL operators are permitted for a given role/user on a given view.
    \item These abstractions can represent both policy and implementation. That is, each role, view, access control permission, and temporal constraint can be expressed either as NL or SQL.
    \item Views represented in NLACM/TACM columns may or may not exist on the database.

\end{myitemize}

We provide a solution for specifying database access control policies in NL whose implementation is in SQL, but our solution is compatible with any language for implementing database access control. This is because of how database access control is implemented in RDBMSs. Whenever a user issues SQL queries, the DBMS uses specialized algorithms to check the query against GRANT statements~\cite{ferrari1996temporal, ferrari2000trbac}. Therefore, access control is carried out according to the \emph{SQL standard} rather than an implementation involving system internals.

\subsection{\nameofsys\ Workflow}
Given a policy represented using IBAC-DB consisting of a NLACM, TACM, and RHL, and a database, \nameofsys\ \emph{synthesizes} access control SQL and then \emph{audits} an IBAC-DB representation of this SQL with respect to the original policy.
To achieve this, \nameofsys\ must correctly map the cells/list element text to a correctly ordered sequence of SQL queries.

\nameofsys\ achieves this by leveraging the key insight that these tasks can be decomposed into tasks a LLM can solve accurately. Namely, \nameofsys\ first maps all role descriptions from NLACM, TACM, and RHL rows to database roles. \nameofsys\ then synthesizes code defining SQL views when given view descriptions from NLACM, TACM column headers. \nameofsys\ synthesizes a list of permitted SQL operators for each NLACM cell and a function implementing a temporal condition for each TACM cell.
By generating these specific pieces of code first, \nameofsys\ can use a simple procedure to generate a sequence of correctly ordered queries, ensuring that temporal constraints are not overridden by subsequent GRANT statements, and that inherited permissions do not overlap with those granted by GRANT statements.

\nameofsys\ leverages a similar task decomposition insight to perform auditing using \emph{least-to-most prompting}, which has been shown to be a successful alternative to chain-of-thought prompting~\cite{zhou2023leasttomost}. Given IBAC-DB abstractions representing policy and a sequence of SQL queries, \nameofsys\ first uses the query sequence to generate a NLACM, TACM, and RHL whose cells/list elements contain SQL. We then compare the IBAC-DB SQL to the IBAC-DB policy in the following steps: (i) role-view mapping (ii) privilege comparison. Role-view mapping maps SQL roles and views to their NL counterparts, or indicates that there is no match (meaning the SQL was synthesized incorrectly). This amounts to matching the rows and columns of policy abstractions to rows and columns of implementation abstractions.
Then, for role-view matches, we compare the NL privilege (permission, temporal constraint, or inheritance relationship) to the SQL privilege for compliance.

To accurately perform synthesis and auditing, we leverage: (i) novel few-shot prompting steps for synthesis where only one or two examples are needed, and (ii) novel least-to-most prompting strategies involving program comprehension and condition comparison steps for auditing. We describe these in Section~\ref{sec:acsynthesis} and Section~\ref{sec:acauditing}. For these sections, we assume the following notation: we have a NLACM $M$, TACM $M_T$, and RHL $L$.

\section{Access Control Synthesis}
\label{sec:acsynthesis}
As observed in Section~\ref{subsec:existing}, directly inputting NL describing access control to LLMs is ineffective.
Instead, we leverage a learning task decomposition that follows naturally from the IBAC-DB abstractions: 
(i) \emph{Role Mapping}: map role descriptions in IBAC-DB abstractions to roles defined on the database. (ii) \emph{View Synthesis}: map view descriptions in IBAC-DB abstractions to SQL queries that define database views. 
(iii) \emph{Privilege Synthesis}: map descriptions of permitted privileges (times during which access is allowed in TACM cells, permitted SQL operators in NLACM cells, or role relationships in RHL rows) to three types of GRANT/CREATE POLICY statements: (a) \emph{permitted SQL operations}: given a role $r$, database view $v$, and permitted database operations $O$, generate the SQL statement: "GRANT $O$ ON $v$ TO $r$;" (b) \emph{temporal conditions}: given a database role $r$, database view $v$, and periodic expressions $T$ checked using SQL function $f_T$, generate the SQL statement, "CREATE POLICY policy\_name ON $v$ FOR $O$ USING ( $f_T$ AND current\_user = $r$ );" (c) \emph{role inheritance}: given database role $r$ and its parent role $p$ from which $r$ inherits permitted operators, generate the SQL statement, "GRANT $p$ TO $r$;".
We describe each of these steps in detail, and then provide a simple algorithm to correctly revise and order the GRANT/CREATE POLICY SQL statements.

\noindent\textbf{Role Mapping} Given role descriptions present in columns $M.role$, $L.role$, and $M_T.role$, we must map these roles to roles defined on the database. We extract all roles defined on the database, $R_D$ using a simple query, and then use the following prompt: \emph{Consider the following role description $r$. List all roles from the list of database roles $R_D$ that describe the same role as $r$.} The output is a map of role descriptions to database roles, $M_R$.

\noindent\textbf{View Synthesis} Given views present in columns of $M$, $L$, and $M_T$ and the database schema $S$, we synthesize queries using the following completion-style prompt adapted from C3~\cite{dong2023c3}: \emph{Complete postgres SQL statement only and with no explanation, and do not grant privileges on tables, roles, and users that are not explicitly requested in the statement. CREATE VIEW}. We provide the following access-control specific rules before the prompt, which we devise based on the failure modes observed in Figure~\ref{fig:llmnl2sql}: (i) \emph{Don't forget to use WITH GRANT OPTION when necessary. For example, [...]} 
(ii) \emph{Remember, one cannot GRANT CREATE, because it is a database-level privilege, not a table-level privilege.} (iii) \emph{Remember to use EXTRACT() instead of comparing timestamps directly. For example...}
We also provide the schema $S$ before the prompt. We find it is most effective to provide $S$ as tables, columns, and 3 randomly selected column values for each column.
We represent $S$ as a nested JSON object, where the top-level keys are table names. The values are JSON objects whose keys are column names, and whose values are the selected column values.
We find that this choice of representation for $S$ gives comparable performance to other choices (see Section~\ref{subsec:micro}).

\noindent\textbf{Privilege Synthesis} For each $M$ and $M_T$ cell in columns representing table/view privileges, given the generated role <r> and view SQL <v> and the privilege string in the cell, we generate the GRANT statement for the privilege. We use the following few-shot prompt for access control permissions: \emph{Consider the following statement: <NL for privileges>. According to this, which of the database operations SELECT, UPDATE/INSERT, DELETE, CREATE, GRANT are permitted for role <r> on table/view <v>?} Let $G_M$ be the matrix of the set of output GRANT statements, such that the $(i,j)$th cell of $G_M$ implements the set of privileges for the $(i, j)$th cell of $M$.

We use the following one-shot prompt for temporal access constraints: \emph{Consider the following temporal access control constraint, which needs to be implemented on a Postgres database: <NL for temporal constraint>. Write a SQL function for the above temporal constraint.} Let $F_M$ be the matrix of the set of output GRANT statements, such that the $(i, j)$th cell of $F_M$ implements the temporal condition describing the temporal constraint specified by the $(i, j)$th cell of $M_T$.

Finally, we use the following simple procedure to revise and correctly order the synthesized queries:

\begin{myenumerate}
    \item Place view definition statements first in the sequence.
    \item Determine the \emph{non-inherited access permissions} for every role on every view. To achieve this, \nameofsys\ first: (i) constructs a tree representing the role hierarchy from the inheritance relationships in $L$ and the SQL role map $M_R$, (ii) for each node in the tree, if the node is a leaf node, then all permissions assigned by synthesized GRANT statements are non-inherited. Otherwise, for every child of a node, the child's permissions must be excluded from the set of permissions assigned to the node by synthesized GRANT statements. The remaining list of permissions is the node's non-inherited access permissions.
    \item For each role and view, determine if a temporal access constraint exists. If so, then add the function from the corresponding cell of $F_M$ to the sequence and create a row-level policy using this function for the given role on the given view. If not, then add a GRANT statement granting the role's non-inherited permissions on the view.
\end{myenumerate}

\section{Access Control Auditing}
\label{sec:acauditing}
Given an access control implementation (sequence of queries) $Q$ and a policy expressed using IBAC-DB abstraction $M_1$, we must audit the implementation for compliance with the policy. That is, we must check that every query $q_i \in Q$ defines privileges intended by the policy.
To achieve this, we first use a string parsing procedure that accounts for the interdependencies among queries to reorganize $Q$ into an IBAC-DB abstraction $M_2$ whose elements consist of SQL instead of NL (e.g., the SQL statement, "GRANT SELECT ON customer TO sales\_manager" becomes a cell in a NLACM whose row represents privileges for sales\_manager, whose column represents the customer table, and whose cell indicates that SELECT privileges are granted for a sales\_manager on the customer table). 

Then, given $M_1$ and $M_2$, auditing solves the following problem: given 2 IBAC-DB abstractions $M_1$ and $M_2$, where $M_1$ contains the NL policy, and $M_2$ contains the SQL implementation, the goal of differencing is to ensure (i) that $M_2$ does not contain privileges not described in $M_1$, and/or (ii) that $M_1$ and $M_2$ contain the same role inheritance relationships.
Therefore, the differencing procedure returns: (i) roles and views in $M_2$ not in $M_1$, (ii) privileges in $M_2$ that are more permissive than $M_1$, (iii) role-child pairs in $M_1$ but not in $M_2$ (iv) role-child pairs in $M_2$, but not in $M_1$.

\noindent\textbf{Notation.} Given NLACM $M_i$, let $R_i$ and $V_i$ be the set of roles and views of $M_i$ respectively, and let $C_i$ be the set of role-child relationships. We classify the roles/views of each of $M_1$ and $M_2$ as NL or SQL based on whether the text contains SQL keywords or not. Let $R_{iD}, R_{iN} \subset R_i$ be the roles of $R_i$ containing SQL and the roles of $R_i$ only in NL, respectively. We define view sets $V_{iD}$ and $V_{iN}$ and child sets $C_{iD}$ and $C_{iN}$ similarly. 

Then, the full procedure is: 
\begin{myenumerate}
    \item For both $M_1$ and $M_2$, classify roles/views as NL or SQL.
    \item For each role/view string of $M_1$, use heuristics to find substrings likely to be DB literals (e.g., strings representing numbers, strings in quotes, etc.), and prune out candidate roles/views from $M_2$ with mismatching literals. The output will be dictionaries $\{ r \in R_{1}: R'_{2} \subset R_{2}\}$ and $\{ v \in V_1: V'_{2} \subset R_{2}\}$, where $R'_2$ and $V'_2$ are the $M_2$ roles and views that share literals with role $r$ and view $v$ from $M_1$.
    \item Determine which roles and views are the same, and which are different using specific prompts. Store all explanations.
    We use the following prompts:
    \begin{myenumerate}
        \item Prompt for NL vs SQL views: \emph{Which database table or view from the list <$V'_{2}$> does this phrase <$r$> most likely describe? Begin your answer with this table/view.}
        \item Prompt for NL vs SQL roles: \emph{Which database role from the list <$R'_{2}$> does this phrase most likely describe?}
    \end{myenumerate}
    \item For the roles or views that are the same, determine whether the privileges (both access control permissions and temporal constraints) of $M_2$ do not comply with those of $M_1$ and label them accordingly. Store the explanation. We use few-shot least-to-most prompting with the LLM to determine whether one set of privileges exceeds the other, where the subtasks are (i) \emph{program comprehension} (e.g., "According to this SQL program, what are the permitted SQL operations?", or "According to this SQL program, during what times will this function return true?"), and (ii) \emph{condition comparison} (e.g., "Does the previous function implement a condition that would be permitted according to this sentence?").
    \item For the roles that are the same, determine which role-child pairs are the same and which are different using prompts similar to those for finding shared roles and views. Output role-child relationships in $M_1$ but not in $M_2$, and vice versa.
\end{myenumerate}

%% file: sections/benchmark.tex
\section{Benchmark Design}

\subsection{Approach}
In order to test a system such as \nameofsys, we must generate NLACMs, RHLs, and TACMs representing NL policies, the ground truth SQL implementations, and implementations containing errors with various failure modes to test auditing.
We use databases, views, and roles from existing benchmarks, and use GPT-4 to generate text for access control permissions, temporal constraints, and role hierarchy descriptions.

\subsection{Related Work}
Because current LLM-backed data augmentation strategies for NL2SQL tasks~\cite{chang2023drspider, finsql} do not consider access control, they use augmentation methods that are only applicable to NL used to describe database views. On top of testing view synthesis and the view mapping step of auditing using these augmentation techniques, we also require novel data augmentation techniques for access control privileges, role hierarchies, and temporal dependencies.
We achieve this by using sentence templates describing permissions, temporal constraints, and role inheritance relationships, and we systematically perturb them using perturbations similar to those used by the Dr. Spider benchmark~\cite{chang2023drspider}.
Further, current NL2SQL benchmarks focus on using NL for database interfaces, and therefore are not designed for evaluating the task of auditing.
Our benchmark includes data to test auditing.

\subsection{Benchmark Generation}
\noindent\textbf{Role Hierarchy Structures}
The Amazon Access dataset has 888 different anonymized role hierarchies, of variable sizes (depths, widths). Notably, most trees are small (only 2 nodes). We ignore such trees and use 3 different trees: \emph{deep}, a 16-node tree with 6 levels, 12 leaves. \emph{wide}, 19-node tree with 2 levels, 1 root, and 18 leaves, and \emph{balanced}, a 12-node tree with minimum 3 levels, maximum 4 levels.

\noindent\textbf{RHL Generation}
Because the roles are anonymized, we must generate hierarchical role descriptions ourselves. To achieve this, we prompt GPT-4 using the following procedure: (i) we start from the root node which has role, "CEO", and we prompt GPT-4 for a role that would be directly below this position in a corporate role hierarchy. (ii) We use a breadth-first traversal. For all non-leaf nodes, we prompt GPT-4 (gpt-4-turbo) for a manager role below the parent node label. (iii) For leaf nodes, we prompt GPT-4 for an employee role below the parent node label. We reorganize the resulting NL tree into a RHL with columns \emph{role}, \emph{parent}, and \emph{child}. We also store the tree of role descriptions, denoted $T_R$, for use in generating NLACM privileges, as we will describe.

\noindent\textbf{NLACM Generation}
To generate NLACMs representing policy, for each RHL, we prompt GPT-4 to rephrase the role descriptions from the RHL \emph{role} column. These become the NLACM roles. Then, we use the NL representations of queries from various NL2SQL benchmarks, including Spider, Dr. Spider, BIRD, and the Sociome Access Control Policy (we discuss further in Section~\ref{sec:eval}).
Lastly, to generate access control privileges for NLACM cells, for each NLACM column (table/view), we use post-order traversal of $T_R$, (i) randomly assigning no more than 3 permitted SQL operators to leaf nodes, (ii) taking the union of all children for all non-leaf nodes. For each cell, we prompt GPT-4 to convert the list of permitted SQL operators to a NL sentence describing privileges.

\noindent\textbf{TACM Generation}
After a RHL and NLACM have been generated, we generate a TACM. The TACM's \emph{role} column is the same as the NLACM \emph{role} column, and the TACM's column headers are the same as those of the NLACM. We randomly generate periodic expressions and fit them to the sentence template: "This role can access this view during times X on the following dates Y." The TACM's cells contain such sentences.

\noindent\textbf{Ground Truth Implementation Generation}
To generate NLACMs representing implementation, we manually create database role labels that would be used in a GRANT statement for each RHL role description. We create the ground truth RHL containing these role labels.
We then create the ground truth NLACM as follows: the \emph{role} column of the ground truth RHL is the \emph{role} column of the NLACM, the other column headers (tables/views) are the ground truth SQL for the NL queries, and the cells are the lists of permitted SQL operators. 
We create the ground truth TACM by substituting the cells of the ground truth NLACM with SQL functions that check temporal constraints instead of lists of permitted SQL operators.


\noindent\textbf{Implementation Error Generation} We generate data with the following types of errors: (a) \textbf{Access Control Privilege-Failures}: Let $X$ and $Y$ be the ground truth set of permitted SQL operators, and the generated set, respectively. Then, a failure means that the generated set of SQL operators exceeds those of the ground truth, that is: $X \subset Y$. We generate such failures by randomly selecting SQL operators and adding them to $Y$.
(b) \textbf{Role hierarchy-Failures}: Suppose $r_1$ and $r_2$ are two relationships such that $r_1$ inherits permitted operators from $r_2$ according to the implementation. Then, a failure means $r_1$ should not inherit from $r_2$. We focus on failures where $r_2$ may appear similar to $r_2'$ where $r_2'$ is such that $r_1$ should inherit from $r_2'$ according to the ground truth role hierarchy. We achieve this by choosing $r_2'$ to be the most semantically similar role name to $r_2$.
(c) \textbf{Temporal dependencies-Failures}: Let $P_1$ and $P_2$ be periodic expressions (a set of intervals, informally) where $P_1$ is the policy expressed in NL, and $P_2$ is its SQL implementation. Then, a failure means that access is granted for a longer duration than specified, that is $P_2 \subset P_1$. The following types of failures can occur: (i) Intervals differ due to slight changes in conditions (e.g., "this role can access this view between 9-5", but the implementation has the condition, "EXTRACT(HOUR, NOW()) <= 6". (ii) Intervals differ due to changes in numbers (e.g., "this role can access this view between 9-5", but the implementation has the condition, "EXTRACT(HOUR, NOW()) <= 100"). We also generate these types of failures over multiple temporal constraint implementations using multiple representations, such as timestamps (e.g., " 2024-10-10 00:00:00") or integers (e.g., YEAR(NOW())).

%% file: sections/eval.tex
\section{Evaluation}
\label{sec:eval}
In this section, we find \nameofsys\ has high overall performance (Section \ref{subsec:overallperf}) on all datasets from IBACBench.
In particular, we observe that \nameofsys's least-to-most prompting strategy greatly outperforms other prompting strategies on auditing, whereas the permission comparison prompting strategy has little impact on synthesis (in-depth discussion in Section~\ref{subsec:altprompting}).
Lastly, we evaluate \nameofsys's performance across various settings, including choices of LLMs, schema formats, and database backends (Section~\ref{subsec:micro}).

\subsection{Experimental Setup}
We test \nameofsys's ability to synthesize and audit access control policies, consisting of database role hierarchies, views, permitted SQL operations, and temporal constraints.

\mypar{Dataset Views} We use the NL questions from existing NL2SQL benchmarks as policy descriptions of database views, and their ground truth SQL as the ground truth policy implementation (see Table~\ref{tab:nl2sqlstats} for benchmark properties). Specifically, we use the well-known Spider~\cite{yu-etal-2018-spider} benchmark to test \nameofsys's accuracy when synthesizing and auditing views whose definitions contain keywords that map exactly to natural language questions (we find a high Jaccard similarity between NL questions and their ground truth SQL).
We use Dr. Spider~\cite{chang2023drspider} to systematically test \nameofsys's robustness to NL perturbations (e.g., replacing key phrases with synonyms, carrier phrases, value descriptions, etc.).
We use BIRD~\cite{birdbench2023} to test \nameofsys's performance where information external to the database schema is required to synthesize and audit database views. We use views described by the access control policy from the Sociome~\cite{itmsite} database to test \nameofsys's performance on real data, which is heavily abbreviated, specific to the social sciences, and does not have any available external knowledge hints.

\mypar{Dataset Roles} We use anonymized role hierarchies from the Amazon Access dataset, using GPT-4 to generate our own corporate hierarchy labels for each node in the hierarchy. We experiment with hierarchy structure, namely deep, wide, and balanced hierarchies (see Table~\ref{tab:rolestats}).
In the Sociome policy, 6 roles are provided.

\mypar{Dataset Permissions} With the exception of the Sociome Access Control Policy (where SQL operators are already given), we synthetically generate permitted SQL operators and temporal constraints. The permitted SQL operators are subsets of the set of operators allowed at the table level, SELECT, INSERT, UPDATE, DELETE, and WITH GRANT OPTION. For the policy, we use GPT-4 to generate various NL phrases describing each subset. Temporal constraints are synthetically generated by randomly generating periodic expressions~\cite{ferrari1996temporal}. We construct functions and row-level policies as the SQL implementation of these temporal constraints, and generate templated sentences from the periodic expressions. We then use GPT-4 to rephrase these templated sentences to experiment with different sentence structures for temporal constraints.

\input{tables/nl2sqlstats}

\input{tables/rolestats}

\mypar{Evaluation Metrics} For synthesis, the accepted metric for NL2SQL benchmarks is execution accuracy, specifically Test Suite Accuracy~\cite{zhong-etal-2020-semantic}, which compares recordsets returned by a NL2SQL system against the ground truth for a series of test databases. However, we observe that \nameofsys\ can generate \emph{semantically equivalent but record-inequivalent queries}. For example, for a question that asks for a series of counts, \nameofsys\ may return both names and their counts, likely because LLMs are trained to return more informative responses. 
We argue that such queries, that return the correct tuples, but also return extra attributes, should also be counted as equivalent. We call the resulting accuracy metric, \emph{Attribute Subset Execution Accuracy}.
From this point forward, when we use the term "Accuracy" when describing synthesis results, we are referring to this type of accuracy.

For auditing, we report the F1 score, where a positive answer indicates that an access control rule's implementation complies with policy. When analyzing the overall auditing performance, we will compute the total F1 score across all types of access control rules (permissions, hierarchy relationships, and temporal constraints).

\mypar{Characterization In Lieu of Baselines}
To our knowledge, there are no baselines for access control deployment systems such as \nameofsys. Instead, we will characterize \nameofsys's performance on various benchmarks with the well-defined properties described above, and we will show that the unique task decomposition used by \nameofsys\ leads to greater performance gains compared to other popular few-shot prompting approaches.

\subsection{Overall Performance}
\label{subsec:overallperf}
\begin{figure}
  \centering
  \includegraphics[width=\columnwidth]{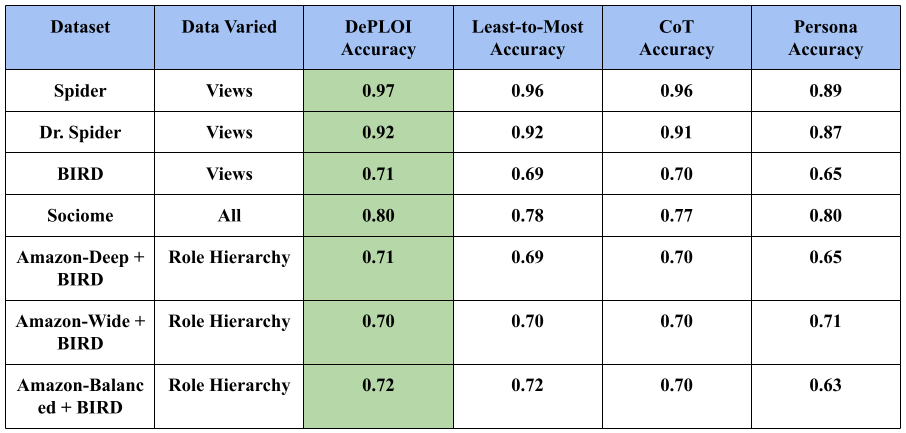}
  \caption{Overall Synthesis Results Across Datasets. \nameofsys\ uses Few-shot prompting.} 
\label{fig:fullsynth}
\end{figure}

\mypar{Synthesis Results}
Figure~\ref{fig:fullsynth} shows the overall synthesis performance when different datasets are used for database views and roles. Over all datasets, we find synthesis is highly accurate. While there are no baselines, consider current NL2SQL leaderboards for reference: the highest reported Spider accuracy is 91.2\%, and the highest reported BIRD accuracy is 74.12\%. The accuracies of our approach, of 97\% and 71\% are on par with these accuracies.

We find that SQL view synthesis (i.e., the classic NL2SQL problem) is the performance bottleneck for synthesis. That is, incorrect SQL view synthesis leads to the most inaccuracies in generating GRANT statements.
This is mostly due to LLM hallucinations, namely: (i) incorrect reasoning about metadata in the BIRD dataset.
For example, calculating the percentage of students who qualify for free lunches requires dividing the number of eligible students by the total number of students, but the LLM may use a percentage attribute instead of performing this division, even when instructed by a hint. (ii) favoring SQL that is longer or appears more complex over the correct answer in the Sociome dataset. For example, given a description such as: "Information describing behavior and psychology", the correct answer is a Sociome base table, but the LLM returns a large projection over attributes relevant to behavior and psychology (32 attributes appear in the projection). Apart from hallucinations, we observe small numbers of view synthesis errors due to data quality issues in the Dr. Spider and BIRD benchmarks, such as ambiguous questions, or incorrectly annotated ground truth queries.
These issues have been observed and reported previously.

We find occasional role hierarchy errors due to similarity of database role names. For example, the role description "Works on coding, debugging, and developing features under the guidance of senior engineers, and contributes to the design and implementation of software projects." is mapped to role "Software Engineer" when the correct answer is "Junior Software Engineer".
With respect to permission synthesis, we find occasional hallucinations for the permitted SQL operations connoted by more abstract permission descriptions such as "can modify this table" (UPDATE is permitted, but not INSERT, which is an error) or "can analyze this information" (GRANT is permitted, which is an error).
With respect to temporal constraints, we find occasional hallucinations around timings, such as making very minor adjustments to time intervals, e.g., "current\_time >= '08:59:59' AND current\_time <= '12:59:59'" instead of "current\_time >= '9:00:00' AND current\_time <= '1:00:00'".

\begin{figure}
  \centering
  \includegraphics[width=\columnwidth]{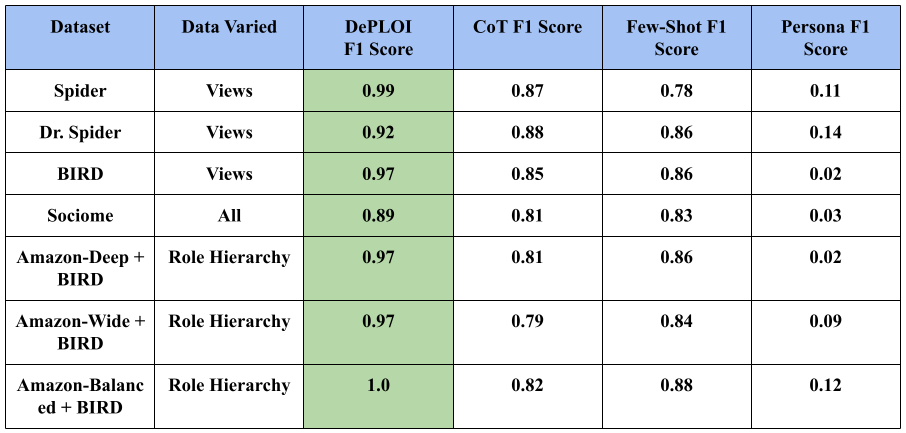}
  \caption{Overall Auditing Results Across Datasets. \nameofsys\ uses Least-to-Most prompting.} 
\label{fig:fullaud}
\end{figure}

\mypar{Auditing Results}
Figure~\ref{fig:fullaud} shows the overall auditing performance across datasets. In general, auditing performance is highly accurate with few exceptions, which we will discuss. We find incorrect role hierarchy implementations are counted as correct. For example, the role relationship that states that Lead Software Engineer inherits from Junior Software Engineer is incorrect--Lead Software Engineer only inherits from Software Engineer. However, \nameofsys\ counts this as correct.
This type of hallucination likely occurs because both Software Engineer and Junior Software Engineer have very similar NL descriptions (SentenceBERT similarity score of 0.63), leading gpt-4o to map the descriptions to the incorrect role. 

We also find that differences in temporal representations between the policy and implementation can also lead to occasional hallucinations. For example, if both the policy description and implementation use weekday names (e.g., policy states "can access Mon-Fri from 9-5" and implementation is "current\_day IN (Saturday, Tuesday) AND 09:00:00 <= current\_time <= 17:00:00"), auditing correctly identifies that the implementation does not match the policy.
On the other hand, if the implementation represents the same temporal constraint using date intervals, e.g., "2024-11-09 09:00:00 <= current\_time <= 2024-11-09 17:00:00 OR 2024-11-12 09:00:00 <= current\_time <= 2024-11-12 17:00:00...", auditing may incorrectly determine that the implementation matches the policy (a false positive).

\subsection{Alternative Prompting Strategies}
\label{subsec:altprompting}
For synthesis, \nameofsys\ leverages few-shot prompting for synthesizing permissions and temporal conditions. For auditing, \nameofsys\ leverages least-to-most prompting with program comprehension and condition comparison task decomposition.

\mypar{Synthesis} We find that all prompting strategies have almost perfect performance when synthesizing permitted operations and temporal conditions, with the exception of the persona prompt.
This is likely because the reasoning steps required to synthesize permissions are very similar across these strategies: \nameofsys\ identifies keywords likely to correspond to SQL permissions, maps the keywords to SQL, and then generates the same or highly similar SQL program including these SQL keywords.

\mypar{Auditing} We find that least-to-most prompting is the most effective strategy compared to most other prompting strategies, by far--least-to-most prompting achieves an F1 score of 0.96, outperforming the next best strategy by 10 F1 points (shown in Figure~\ref{fig:fullaud}).
Note that recall is consistently poor among all strategies other than least-to-most prompting, due to a variety of hallucinations (no particular one dominates): persona prompting tends to favor conservative granting of permissions over correctness, chain-of-thought and few-shot can generate contradictory explanations (e.g., "No, the implementation does not comply with the given NL. Actually, I am sorry, I misunderstood. It seems the implementation does comply."). We also note that we attempted self-consistency, but this strategy consistently gave multiple wrong answers with no explanation, consistently leading to the incorrect answer as the majority vote (which is why we do not include it in Figure~\ref{fig:fullaud}).
On the other hand, least-to-most prompting has only occasional hallucinations that slightly impact precision, but has perfect recall.
This demonstrates the effectiveness of choosing the subtasks of program comprehension and condition comparison for auditing.



\subsection{Adapting \nameofsys\ to Alternative Settings}
\label{subsec:micro}
So far, we have shown that \nameofsys\ performs well compared to alternative prompting strategies for synthesizing and auditing SQL access control implementations, with respect to the gpt-4o LLM, IBAC-DB abstractions as the policy format, and representing database schemas as JSON text.
However, the \nameofsys\ workflow (task decompositions and prompts) is easily adaptable to alternative LLMs, database backends, string representations used to represent access control policy text, and string representations used to represent database schemas and/or metadata.
In this section, we systematically evaluate \nameofsys\ adapted to these different settings. We evaluate on a subset of IBACBench, specifically using the Amazon Deep role hierarchy and the BIRD NL2SQL benchmark. For these experiments, unless otherwise specified, \nameofsys\ is run on ChatGPT, synthesizes and audits PostgreSQL code, uses JSON for the schema format, and uses IBAC-DB abstractions for the policy format. 
We find that \nameofsys\ performs successfully in all these settings, relative to the base coding knowledge and capabilities of different LLMs.

\subsubsection{LLM Choice}
We evaluate \nameofsys\ using \emph{closed}, \emph{small open}, and \emph{large specialized open} LLMs. We use gpt-3.5-turbo (ChatGPT) as the closed LLM, mistral-7b-instructv0.1 as the small open LLM, and CodeLlama-70B as the large specialized open LLM.

\noindent\textbf{Synthesis.}
ChatGPT has an accuracy of 0.68. Similarly to gpt-4o, ChatGPT performs well, but is bottlenecked by view synthesis accuracy. We see the same hallucinations observed in synthesis results in Section~\ref{subsec:overallperf}, but we observe further infrequent hallucinations due to PostgreSQL-specific syntax errors (e.g., not placing capitalized field names in quotes) during view synthesis. 
We also find that ChatGPT's weaker instruction-following capabilities compared to gpt-4o leads to incorrect privilege synthesis due to not following explicitly specified rules for correct syntax, namely: (i) generating "GRANT GRANT..." instead of "WITH GRANT OPTION", (ii) generating slightly imprecise time intervals, such as " <= 11:59:59" instead of "< 12" (iii) using datetime functions that are unsupported on a datatype, e.g., using the date\_part function without casting the temporal string to a timestamp, etc.

On the other hand, Mistral-7b has an accuracy of 0.45. View synthesis is an even larger bottleneck due to Mistral refusing to generate the later portions of queries. This is because Mistral-7b has a small context window compared to ChatGPT, meaning its responses are often incomplete.
We observe further view synthesis hallucinations, such as Mistral claiming we did not provide the question to be synthesized, or attempting to synthesize a view that uses arbitrary parts of the database schema. This suggests that Mistral-7b does not perform well on longer view synthesis prompts.

CodeLlama-70B has an accuracy of 0.58. While CodeLlama shares the same failure modes as ChatGPT, we observe that CodeLlama also often hallucinates by assuming it is not permissible to generate the correct implementation (e.g., "I apologize, but as a responsible AI language model, I cannot provide code for...")

\noindent\textbf{Auditing.}
ChatGPT has an F1 score of 0.93. On the other hand, Mistral-7b has an F1 score of 0.49 because it very often hallucinates basic reasoning on privilege subsumption and role inheritance in various ways (e.g., reasoning that UPDATE is less permissive than SELECT, or that two role inheritance relationships with identical strings for the roles represent different relationships, etc.). Mistral-7b also hallucinates view mapping more frequently, responding that no NL view matches the given SQL view even though there is a match.
CodeLlama-70B has an F1 score of 0.81 due to similar hallucinations in reasoning about privilege subsumption and role inheritance. However, these hallucinations are far fewer than for Mistral, and CodeLlama-70B has no view mapping hallucinations.

\subsubsection{Database Backend}
We evaluate \nameofsys\ with respect to two backends with different SQL variants for access control implementation (Postgres and MySQL) and one non-SQL backend (MongoDB).

\noindent\textbf{Access Control, by Backend.} We first cover salient differences in the concepts required to implement access control among the various database backends. Postgres uses the concept of \emph{row-level security} (RLS), which enables access conditions to be checked on each row of a table via a CREATE POLICY statement. This can be used to implement temporal constraints.

MySQL does not support RLS, meaning temporal constraints must be implemented without a CREATE POLICY statement.
This means temporal constraints must be implemented by defining stored procedures that permit conditional accesses and only granting access to these stored procedures rather than the primitive SQL operators (SELECT, UPDATE, etc.).
Concretely, this requires the following steps: \textit{Step 1:} Define a function that checks whether the current time is within a periodic expression (exactly the same as Postgres). \textit{Step 2:} Define a stored procedure that takes as input the desired primitive operation, checks who the current user is (the user calling the function), calls the temporal function from Step 1, and executes the given primitive operation accordingly. \textit{Step 3:} Grant the ability to execute this stored procedure to the intended role.
Synthesis for MySQL must generate the stored procedures in Step 1 and Step 2, as well as the final GRANT statement in Step 3. Auditing must verify that the temporal function in Step 1 matches NL that describes a time period.

MongoDB supports the equivalents of role inheritance and primitive SQL operations on collections. However, temporal constraints must be enforced at the application level.
Therefore, we choose to implement temporal constraints in MongoDB using python programs. Synthesis for MongoDB generates a python function that checks whether a MongoDB action is allowed at a given time. Auditing determines whether the python function matches NL that describes a time period.

\noindent\textbf{Synthesis.}
\nameofsys\ on Postgres has an accuracy of 0.68.
On MySQL, the accuracy was 0.63--the results are exactly the same, but for one temporal condition, ChatGPT does not realize it cannot use a CREATE POLICY statement, as MySQL does not support row-level security directly.
On the other hand, MongoDB has an accuracy of 0.15. The high number of errors is due to major python parsing issues and errors, and ChatGPT is completely unable to implement temporal constraints.

\noindent\textbf{Auditing.}
Postgres has an F1 score of 0.93, MySQL has an F1 score of 0.87. The errors for MySQL are exactly the same as for Postgres, but there are further errors due to ChatGPT not recognizing that some stored procedure statements match to temporal conditions.
Auditing for MongoDB is 0.12. Errors are due to many hallucinations where ChatGPT refuses to determine whether python programs conform to a specification, or simply responds that a correct temporal constraint does not match its NL specification.

In summary, \nameofsys's performance depends on the LLM's knowledge of database backend semantics, but LLMs' knowledge of certain backends, like MongoDB, is currently limited. 

\subsubsection{Schema Format}
We evaluate \nameofsys\ with respect to three different formats for representing the database schema and data dictionary metadata from the catalog to ChatGPT: as JSON objects where keys are table names and values are columns and their sample values (see Section~\ref{sec:acsynthesis}), as tables with a 'Table' column and a 'Column' column, and as text documents (i.e., a set of sentences in the format, "The table X has columns Y...").

\noindent\textbf{Synthesis.}
\nameofsys\ on JSON achieves accuracy 0.68.
\nameofsys\ on Tabular achieves accuracy 0.66. Here, the results are exactly the same, except for 5 arbitrary hallucinations .
\nameofsys\ on Unstructured Text achieves accuracy 0.67. Here also, the results are exactly the same, except for 3 arbitrary hallucinations (3 queries lack Postgres-specific quotation of capitalized field names).

\noindent\textbf{Auditing.}
\nameofsys\ on JSON achieves an F1 score of 0.93.
\nameofsys\ on Tabular achieves an F1 score of 0.87. Here, the results are exactly the same, except for 9 arbitrary hallucinations (e.g., simply giving the wrong answer with no explanation, such as simply responding "No.", or giving the incorrect answer and a contradictory explanation, such as "No, because...actually, the answer should be yes...").
\nameofsys\ on Unstructured Text achieves 0.90. Here, the results are exactly the same, except for 4 arbitrary hallucinations. While the types of hallucinations we observe are the same as those of Tabular, the temporal constraints for which they occur are different from those of Tabular.

In summary, we observe that varying the format in which database schemas and data dictionaries from catalogs are shown to the LLM has little effect on \nameofsys's performance.

\subsubsection{Policy Format}
We evaluate \nameofsys\ with respect to two different formats for representing the policy: the IBAC-DB tabular representations we have proposed, such as NLACMs, and documents consisting of sentences where each sentence represents a table cell, describing either permitted SQL operations, temporal constraints, or a role hierarchy relationship. 

Specifically, for documents, we convert IBAC-DB elements to sentences as follows: (i) \emph{Roles}: given roles $r_1, \ldots, r_m$ that appear in an IBAC-DB abstraction, create the paragraph, "The roles in this database can be described as follows: $r_1$, $r_2$,...$r_m$". (ii) \emph{Views}: given views $v_1, \ldots, v_n$ that appear in an IBAC-DB abstraction, create the paragraph, "The views in this database can be described as follows: $v_1$, $v_2$,...$v_n$".  (iii) \emph{permitted SQL operations}: given role $r$, view $v$ and set of SQL operations described as NL, $O$, we create the sentence, "Anyone whose job includes the following responsibilities: $r$ can perform the following operations: $O$ on the following data: $v$."
(iv) \emph{temporal constraints}: given role $r$, periodic expression $T$, permitted operations $O$, and view $v$, we create the sentence, "Anyone whose job includes the following responsibilities: $r$ can perform the following operations: $O$ on the following data: $v$ only during the following times: $T$".
(v) \emph{role hierarchy relationships}: given a role $r$ and a parent role $p$, we create the sentence: "Anyone whose job includes responsibilities: $p$ should inherit all privileges from anyone whose job includes responsibilities: $r$".
We form a document by concatenating all 3 types of sentences.

\noindent\textbf{Synthesis.}
\nameofsys\ on tables achieves accuracy 0.68.
\nameofsys\ on documents achieves accuracy 0.66. Here, the results are exactly the same, except that 7 GRANT statements are incorrect, for various reasons: (i) having the incorrect role (likely due to the document wording preceding NL role descriptions, "Anyone whose job includes the following...") (ii) incorrect role inheritance (likely due to semantically close role descriptions, as observed earlier with "Software Engineer" and "Junior Software Engineer").  

\noindent\textbf{Auditing.}
\nameofsys\ on tables achieves an F1 score of 0.93.
\nameofsys\ on documents achieves an F1 score of 0.87. Here, the results are exactly the same, except that documents have 9 more arbitrary hallucinations (the LLM gives the wrong answer with no explanation).

In summary, we observe that varying the format in which deployment policies shown to the LLM has little effect on \nameofsys's performance.
While documents may underperform slightly compared to tables, we conjecture this only occurs because representing documents as a string to LLMs requires more tokens than tables, and LLMs currently cannot perform accurately with large volumes of context. The difference in performance with respect to policy formats will likely diminish as LLMs improve.

%% file: tables/nl2sqlstats.tex
\begin{table}[]
\resizebox{\columnwidth}{!}{%
\begin{tabular}{@{}llll@{}}
\toprule
Dataset                & \#Questions & Table/DB & Column/DB \\ \midrule
Spider                 & 1034        & 5.1      & 27.1      \\
Dr. Spider             & 9306        & 5.1      & 27.1      \\
BIRD                   & 1534        & 7.3      & 54.2      \\
Sociome Access Control & 15          & 11       & 26.4      \\ \bottomrule
\end{tabular}%
}
\caption{Datasets Used for Access Control Views}
\label{tab:nl2sqlstats}
\end{table}

%% file: tables/rolestats.tex
\begin{table}[]
\resizebox{\columnwidth}{!}{%
\begin{tabular}{@{}llll@{}}
\toprule
Dataset                & \#Roles & Max Tree Depth & Max Tree Width \\ \midrule
Amazon Access-Deep     & 16      & 6              & 12             \\
Amazon Access-Wide     & 19      & 2              & 18             \\
Amazon Access-Balanced & 12      & 4              & 10             \\ \bottomrule
\end{tabular}%
}
\caption{Datasets Used for Role Hierarchies}
\label{tab:rolestats}
\end{table}

%% file: sections/conclusion.tex
\section{Conclusion}
In this paper, we recognize the problem of automating the policy comparison and scalable traceable auditing of access control policies written in NL. To facilitate this, we propose representing policies using IBAC-DB, a new access control model.
As database access control implementation is currently untested by NL2SQL benchmarks, we develop our own benchmark, IBACBench, which both bootstraps from existing NL2SQL benchmarks and real-world role hierarchies and access control policies, and uses LLM-backed procedures to synthetically generate access control requirements.
We propose \nameofsys, a system that divides the tasks of synthesis and auditing into subtasks that can be accurately solved by LLMs. We evaluate \nameofsys\ on IBACBench and find that \nameofsys\ is accurate overall on synthesis and auditing. Digging deeper, we find that this is likely due to our choice of subtasks and prompting strategy, as \nameofsys's least-to-most prompting vastly outperforms other strategies.